# Maximum-Likelihood Estimation of Glandular Fraction for Mammography and its Effect on Microcalcification Detection

Bryce J. Smith[1,2], Joyoni Dey[1*] , Lacey Medlock[1], David Solis[2], and Krystal Kirby[2]

[1] Department of Physics and Astronomy, Louisiana State University, Baton Rouge, USA
[2] Physics Department, Mary Bird Perkins Cancer Center, Baton Rouge, USA
*deyj@lsu.edu (corresponding author)

**Abstract**
**Objective:** Breast tissue is mainly a mixture of adipose and fibro-glandular tissue. Cancer risk and risk of undetected breast cancer increases with the amount of glandular tissue in the breast. Therefore, radiologists must report the total volume glandular fraction or a BI-RADS classification in screening and diagnostic mammography. A Maximum Likelihood algorithm is shown to estimate the pixel-wise glandular fraction from mammographic images. The pixel-wise glandular fraction provides information that helps localize dense tissue. The total volume glandular fraction can be calculated from pixel-wise glandular fraction. The algorithm was implemented for images acquired with an anti-scatter grid, and without using the anti-scatter grid but followed by software scatter removal. The work also studied if presenting the pixel-wise glandular fraction image alongside the usual mammographic image has the potential to improve the contrast-to-noise ratio on micro-calcifications in the breast.
**Approach:** The algorithms are implemented and evaluated with TOPAS Geant4 generated images with known glandular fractions. These images are also taken with and without microcalcifications present to study the effects of glandular fraction estimation on microcalcification detection. We then applied the algorithm to clinical DICOM images with and without microcalcifications.
**Results:** For the TOPAS simulated images, the glandular fraction was estimated with a root mean squared error of 3.2% and 2.5% for the without and with anti-scatter grid cases. Average absolute errors were (3.7 +/- 2.4)% and (3.6 +/- 0.9)%, respectively. Results from DICOM clinical images (where the true glandular fraction is unknown) show that the algorithm gives a glandular fraction within the average range expected from the literature. For microcalcification detection, the contrast-to-noise ratio improved by 17.5-548% in DICOM images and 5.1-88% in TOPAS images.
**Conclusion:** We show a method of accurate estimation of pixel-wise glandular fraction image, providing localization information of breast density. The glandular fraction images also showed an improvement in contrast to noise ratio for detecting microcalcifications, a risk factor in breast cancer.

## 1. INTRODUCTION

Breast volume glandular fraction (VGF) is a measure of the extent of glandular tissue present in the breast. It is calculated as the ratio of volume of fibro-glandular tissue to adipose tissue in the breast. Higher breast density is considered risk factor for cancer [1]. It also presents increased likelihood of undetected tumors. Radiologists need an estimated Volume Glandular Fraction (VGF) for screening and diagnostic mammography because mammographic density can vary significantly among patients and change over time [1-3]. To assess breast composition, the American College of Radiology developed the Breast Imaging Reporting and Data System (BI-RADS), which categorizes breast composition into four categories: 1 (predominantly fatty), 2 (scattered fibro-glandular densities), 3 (heterogeneously dense), and 4 (extremely dense). BI-RADS is used to assess breast cancer risk, but its reliability and validity are still debated [4]. A limitation of BI-RADS is that it reports glandular fractions using descriptive text, lacking a quantitative value [5]. Most importantly, the United States FDA has recently standardized the breast density information provided to patients by mammogram facilities. Physicians are required to provide a fibro-glandular density estimate and communicate to patients the impact of breast density on future diagnoses and imaging [6].

Convolutional neural networks (CNNs) have recently been applied to predict breast density, showing comparable performance to human readers [7-8]. For instance, the results of Deep-LIBRA (Maghsoudhi et al, *Med. Imag. Anal*, 2021) [7] and that of an expert reader were highly correlated (Spearman correlation coefficient 0.9). In a 2022 study by Magni et al. in *Radiology* [8], CNN achieved 89.3% accuracy in distinguishing between BI-RADS non-dense and BI-RADS dense-breast categories. While AI-based methods are efficient and improve hospital workflow, reduce workload of Radiologists, they still rely on the qualitative BI-RADS categorization system, which may introduce a degree of subjectivity.

Physics-based studies have been conducted in the past to get a quantitative value to complement BI-RADS such as van Engeland et al. [9], where FFDMs with scatter reduction techniques were studied. However, there remains some uncertainty in the best approach when it comes to quantitative estimation of breast density [10]. There are two main types of physics models that have been studied when it comes to glandular fraction estimation: so called "absolute physics" and "relative physics" models. For the "absolute physics" case, some raw acquisition information is needed. The model of study in recent literature has been the relative physics model since it drastically reduces the need for accurate imaging physics data, which has led to the development of certain commercial algorithms that are in use today [11].

In this study, we will estimate the glandular fraction (GF), as a thickness ratio, the fibro-glandular tissue thickness to the breast tissue thickness, pixel by pixel from the full-field digital mammography (FFDM) image. The advantage of providing a pixel-wise GF image is that it provides *localization information* of higher density tissue. We can derive the VGF from pixel-wise GF.

We will be optimizing an "absolute" physics model, which considers available raw detector images before post-processing, in order to most accurately estimate breast density.

Our work is most similar to the work done by Highnam et al. [12] with key differences. The Highnam et al. method uses a mathematical formulation, but without optimization and has some estimated errors of glandular thickness more than 1 cm for the worst cases. Our model uses a physics-based statistical model, iteratively optimizing the Poisson likelihood function.

Since the large amount of scatter in the mammography energy range tends to degrade image quality, scatter reduction techniques will improve image quality and aid in the estimation of the glandular fraction (GF). Our model incorporates scatter as means to reduce scatter from images. The study will consider two different scatter reduction techniques. One is scatter reduction via standard hardware, such as anti-scatter grids (ASG) and second is via software-based scatter correction algorithms. Anti-scatter grids can eliminate up to 85% of incident scatter but have some drawbacks. The ASG absorbs primary X-rays requiring increased exposure to maintain same fluence at the detector, which results in increased dose to the object. They also require physically demanding alignment, since even slight misalignments can cause artifacts [13]. The use of software-based algorithms has the advantage of no dose increase and no potential artifacts that can arise from a physical grid.

Microcalcifications represent a known risk factor in breast cancer [14]. Therefore, we also undertook a study to investigate whether our estimated pixel-wise GF image could enhance the clinical detection of microcalcifications, particularly in challenging cases.

In what follows we developed a two-step algorithm, first to remove scatter and then estimate glandular fraction: **(1)** a maximum log-likelihood expectation algorithm based on Poisson statistics to remove scatter algorithmically (ML-ScRmv) or via an equivalent of an anti-scatter grid (ASG-ScRmv) and then **(2)** estimate the glandular fraction (ML-GF). The poly-energetic spectrum was approximated by an equivalent mono-energetic spectrum and the concept of effective attenuation [15,16]. The algorithm was tested with TOPAS Monte Carlo generated images. The error due to this approximation was evaluated. Microcalcification contrast to ratio was also assessed in the GF image versus the original mammography image. Finally, we applied the algorithm to a few clinically obtained DICOM images for a patient for pixelwise GF estimation and microcalcification contrast to noise assessment for preliminary clinical testing.

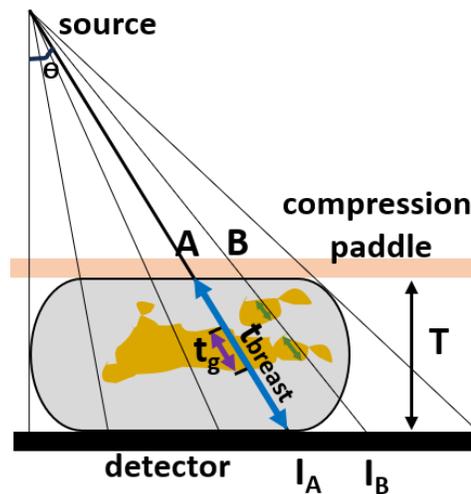

*Figure 1.* Schematic for ray-by-ray (detector pixel to source) fibro-glandular tissue estimation. **T** is the compressed thickness. The slanted thickness $t_{breast}$ (blue) is the intercept of Ray A through the entire breast and $t_g$ (purple) is the length through the glandular part. The Glandular Fraction GF is $t_g/t_{breast}$ which we wish to estimate for detector pixel corresponding to ray A.

## 2. METHODS

Our method can be explained briefly as follows. From a single projection, knowing the thickness of the compressed breast T (from DICOM-header information) and counts on the detector before and after the object placement, we can estimate the net attenuation coefficient at each pixel. This estimation will include models for x-ray scattering to eliminate scatter. Then, using multi-tissue model, and with known values of attenuation coefficients of fibro-glandular, adipose, and skin, and estimating/assuming skin layer thickness, the glandular fraction (the glandular tissue length ($t_g$) as a fraction of the total intercept of the ray through breast, $t_{breast}$) can be estimated ray-by-ray (Fig. 1) that is, at each detector pixel. The paddle thickness is also accounted for. For the method to work for a polychromatic spectrum we applied the concept of effective energy.

*2.1 Iterative Maximum Likelihood (ML) algorithms to estimate GF*

The method involves two steps. In the first step, given a full field digital mammography (FFDM) image, a scatter correction is applied via an iterative ML algorithm. This step is not required if an anti-scatter grid was present during acquisition. In the second step, the GF is estimated by employing a second iterative ML algorithm.

*Step 1. ML for scatter correction. (**ML-ScRmv**)*

Similar to the maximum likelihood method we implement to estimate the thickness in a previous literature [17], here we estimate the $\mu$, linear attenuation coefficient, assuming that the compressed breast thickness is known from the paddle information in the DICOM header of the FFDMs.

We modeled the scatter of the TOPAS setup by using the following equation, were $f(\mu)$ is the likelihood function assuming that the measured counts $I_m$ with object in place is Poisson distributed.

$$f(\mu) = (I_m)\ln(I_e) - I_e \qquad (1)$$

The term $I_e$ is the expected value of the count, for a given $\mu$ and $I_0$.

$$I_e = I_0 e^{-\mu t} + Sc(t). \qquad (2)$$

where $I_0$ is the counts without object in place, $Sc(t)$ is the modeled scatter as a function of thickness and $t$ is the total thickness of the object traversed by the x-ray, where the total path length takes into account the adipose tissue, glandular tissue, skin, and Lexan (compression paddles)

Considering Eq. 1 and 2, the $\mu$ was updated iteratively through applying Newton Raphson's root finding of the first derivative, $f'(\mu)$, to maximize $f(\mu)$ shown in (3) below

$$\mu^{n+1} = \mu^n - \frac{f'(\mu)}{f''(\mu)}, \qquad (3)$$

where $f'(\mu)$ and $f''(\mu)$ are the first and second derivatives of (1) with respect to $\mu$. This gave us an updated scatter-free count image $I_q$ as shown in (4). This was used as input in next step.

$$I_q = I_0 e^{-\mu t} \qquad (4)$$

*Step 2. ML for GF estimation (**ML-GF**)*

In Step 2, a second ML algorithm was used to calculate the GFs. This optimization iteratively maximizes the Poisson log-likelihood of X-ray counts without scatter $I_q$ as shown in Eq. 4 using Newton Raphson method. This time, we write the same log likelihood equation with scatter-free image $I_q$. We estimated the GF (glandular fraction, $m_1$) of the phantom at each detector pixel encountered by the X-ray by using the following Poisson log-likelihood function

$$g(m_1) = (I_q)\ln(I_{e1}) - I_{e1}, \qquad (5)$$

where the expected model was updated to $I_{e1} = I_0 e^{-\mu_1 t}$ as the scatter-less or scatter-corrected image.

The total model $\mu_1$ is split into four variables to account for the different materials that the x-rays traverse through: paddle material, assumed Lexan ($\mu_l$), skin ($\mu_s$), adipose tissue ($\mu_a$), and glandular tissue ($\mu_g$). This leads to the following equation

$$\mu_1 = \mu_l * j + \mu_s * k + \mu_a * (1 - k - m_1 - j) + \mu_g * m_1, \qquad (6)$$

where $m_1$ is the ratio of the glandular tissue thickness to the total path length. It is important to note that $I_q$ accounts for attenuation through the breast and paddle. So, at this stage for $m_1$, the total path includes the breast and compression paddles. We will perform a correction for this to get the true glandular fraction. The fraction j is that of paddle (of material Lexan) in path of x-rays, and k is the fraction of skin in path of x-rays. These fractions are the thickness of each component divided by the total intercept of the ray through the breast and paddle. The lengths through Lexan and skin, j, k, are assumed known, assumed or estimated. Here we assumed that the paddle is 2.5 mm in thickness and a skin thickness of 1.45 mm was chosen [3].

The glandular ratio, $m_1$, then can be estimated via the Newton Raphson updates as follows

$$m_1^{n+1} = m_1^n - \frac{g'(m_1)}{g''(m_1)}, \qquad (7)$$

where $g'(m_1)$ and $g''(m_1)$ are the first and second derivatives of (5) with respect to $m_1$. To eliminate the compression paddles from the fraction to leave just the length through breast, we multiplied the glandular fraction $m_1$ by the total setup thickness, $t_{total}$. This is the net glandular thickness distributed along the ray. Dividing this by the thickness through the breast $t_{breast}$ was then used to get final glandular fraction, $m$.

$$m = \frac{m_1 * t_{total}}{t_{breast}} \qquad (8)$$

This removes the effect of paddle thickness from the $m_1$-estimate. We apply this method on finding the glandular fraction GF on compressed simulated phantoms made up of primarily glandular and adipose tissue and clinical datasets.

Calculation of Volume Glandular Fraction (VGF) can be done by finding the lengths of fibro-glandular tissue for each pixel $m * t_{breast}$ and summing over all pixels and then dividing the lengths $t_{breast}$ summed over all pixels.

We evaluated the algorithm using TOPAS, which wraps and extends the GEANT4 Monte Carlo program [18-19]. We used it to generate images of a semicylindrical object resembling a breast phantom with a heterogenous mixture of glandular and adipose tissue. In TOPAS simulations, we could extract the scatter-to-primary ratio (SPR) from individual simulations by separating scatter from primaries post-simulation. This allows us to create a pixel-by-pixel SPR map for use to model scatter, Sc(t). Anti-scatter-grid case was indirectly simulated. The simulations are explained in the next two sections.

*2 Evaluation with TOPAS Monte Carlo simulations*

TOPAS Monte Carlo (version 3.8.1) software [18,19] was used to generate realistic X-ray images for a mammography system. We used physics list Option4, which includes photoelectric effect, coherent and incoherent scattering. The detector is a Cesium Iodide detector with dimensions 20 cm x 15 cm x 6 mm. The source to detector distance is 70cm and the source to distal side of object distance is 68.5cm. This corresponds to a 1.5cm air gap, which is a similar setup described in Boone et al. [20]. The anode source was a 28kV poly-energetic W/Al (tungsten anode with 0.7mm aluminum filter) spectrum. The spectrum energy domain was considered from 14-28 keV, peaking around 20 keV. Fig. 2(a) and 2(b) show the setup without and with the x-rays turned on, with white wireframe object consisting of adipose tissue while the pink wireframe object represents the glandular tissue, which is the fraction, $m$, that is being estimated. We calculated the $\mu_a$ to be values given from NIST XCOM database after constructing the mixture based on the materials listed in the GEANT4 materials database for adipose tissue. For glandular tissue, we used $\mu_g$ to be values from NIST XCOM after constructing the mixture based off the materials listed by Hammerstein et al [21]. Additionally, the Lexan compression paddles (2.5mm thick), a 1.45mm skin layer, and a water bath were included to simulate the body. For each energy bin, 120million counts were simulated, ensuring sufficient statistical accuracy [22]. For our GF estimation purposes, the detector was binned to 1mm.

To evaluate the algorithm, we used breast thicknesses in 1 cm increments from 3 to 5 cm, with a diameter of 15cm and chest wall to nipple distance of 7cm [23], and glandular fractions of 20%, 30%, and 50%. The glandular tissue was concentrated within a smaller half-cylindrical volume (pink volume in Fig. 2(a)), surrounded by the larger half-cylindrical adipose volume (as shown in Fig. 2(a)). These numbers were chosen to simulate the average range of glandular breast tissue.

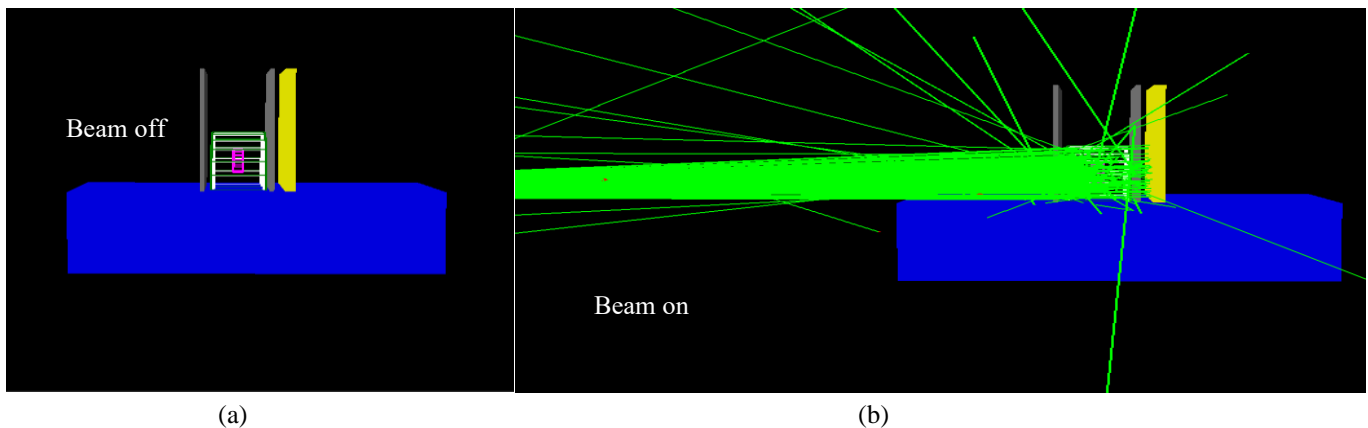

(a)          (b)

*Figure 2. (a) Mammography setup as shown in TOPAS. Objects from left to right are water bath to simulate body (blue), upstream compression paddle (gray), skin layer (green), adipose tissue (white), glandular tissue (pink), downstream compression paddle (gray), and detector (yellow). (b) Mammography setup as shown in TOPAS with beam turned on.*

Fig. 3(a) and 3(b) show the input $I_0$ and $I_m$ images used in the ML algorithm to estimate the glandular fraction for an example case, while Fig. 3(c) shows the outcome of the ML algorithm for this example.

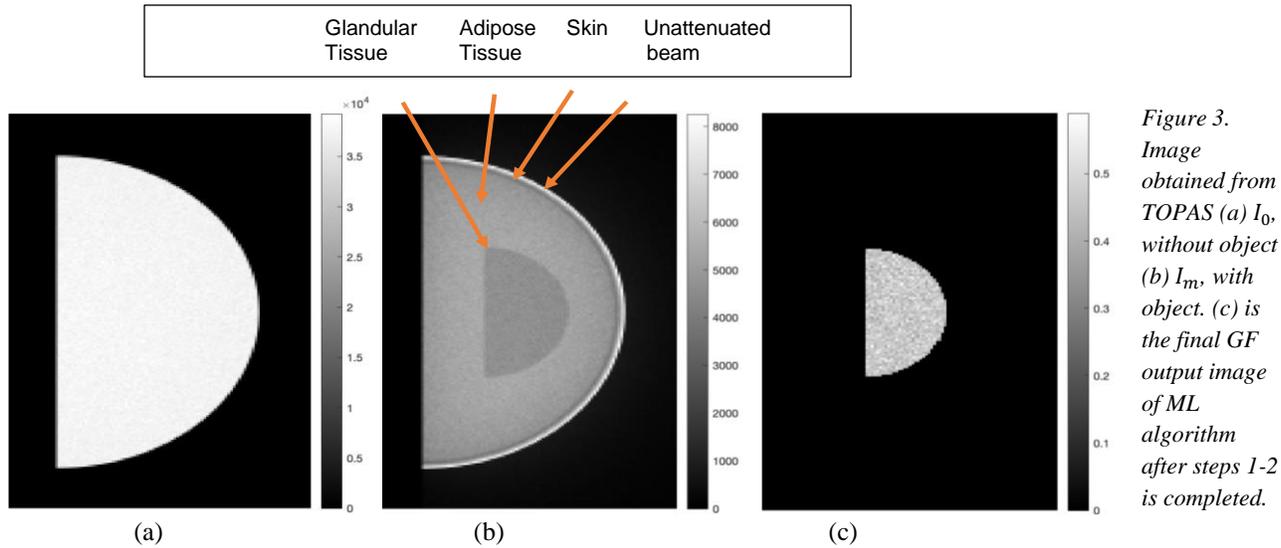

Figure 3. Image obtained from TOPAS (a) $I_0$, without object (b) $I_m$, with object. (c) is the final GF output image of ML algorithm after steps 1-2 is completed.

*2.3 Anti-Scatter Grid scatter removal simulations in TOPAS (ASG-ScRmv)*

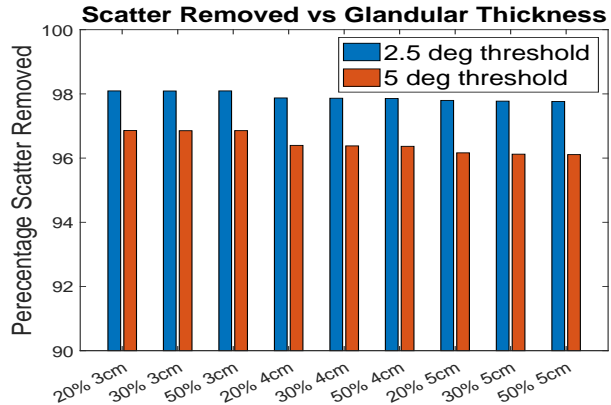

*Figure 4. Percentage scatter removed for each setup for 2.5° & 5°*

We wish to evaluate our GF estimation for cases where the software-based scatter correction Eq. 1-3 will be used as well as for cases where an ti-scatter-grid (ASG) is used in acquisitions. For the anti-scatter grid simulation, instead of a direct grid, a flexible method was implemented in TOPAS to achieve the effects of clinical mammography anti-scatter grid. The flexible method could be potentially used to test effects of different types of anti-scatter grid by controlling the levels of remnant scatter. The method is as follows.

From the TOPAS histories, the direction cosines of the momentum vector with respect to x and the y are known for each individual history counted by the detector. Using these x and y component of momentum vectors, the angle that each ray hit the detector are calculated. Any angle larger than a certain value can be estimated to be scatter and removed from the image. Previous literature shows the mean scatter angle for mammography projections is 5-20 degrees [24]. Two threshold angles 2.5 deg and 5 deg were tested in this work and the scatter removal thereof was estimated. This was estimated from the counts in the region outside the object which is scattered photons. The amount of scatter was estimated as follows. The x-rays detected well beyond the radius of the (outer) semicylinder object and outside of beam were considered scattered rays. The number of counts detected beyond the object (where only scattered photons will reach) were compared before and after thresholding by angle to determine how much scatter was removed. Fig. 4 shows the removal by 2.5 and 5 degrees. The threshold of 2.5 deg was chosen since it was estimated that this angle removed about 98% of the scatter compared, as shown in Fig. 4. To simulate real clinical conditions, 25% of the removed scatter was then added back since a typical anti-scatter grid only removes 75%-85% of the scatter [13]. This gave an estimated "scatter free" $I_m$ akin to what would be obtained with ASG and then used as an input to Eq. 5 for the GF estimate via ML.

*2.4 Polychromatic Spectrum*

So far in Eq. 1-4 the polychromatic nature of the spectrum was not considered. In this work we used attenuation coefficients at an effective energy and estimated the errors thereof using TOPAS. The total equivalent attenuation is calculated first for the given thickness (of object and paddle) for the detector images (obtained with a polychromatic spectrum). The effective energy for this equivalent total attenuation ($\mu$) is found using linear interpolation among our known spectrum range. We estimate the adipose tissue ($\mu_a$), and glandular tissue ($\mu_g$) for the algorithm at this effective energy. This method is then evaluated. The percent error of images objected with polychromatic attenuation versus effective energy attenuation is tested in TOPAS. For all cases, the effective energy attenuation yielded less than 1% error as shown in the Results sections (Table I). Other validations such as poly-energetic versus monoenergetic attenuations for different glandular thicknesses are also shown and discussed in Results section.

*2.5 Addition of Microcalcification to Monte Carlo Images*

Microcalcifications are considered to be robust markers for breast cancer, where 30-50% of non-palpable cancers are detected by microcalcifications revealed during mammography [25]. This makes detecting them early an important feature of mammography. Detection of type II (hydroxyapatite) calcifications is crucial since those are associated with malignant lesions [26]. Microcalcifications were simulated in the TOPAS images by inserting calcium deposits in two different spots of the glandular portion of the 3D object before obtaining the projection data. These deposits are in a cylindrical shape with a diameter of 1 mm and a height of 0.5mm.

*2.6 Clinical Images*

*A. Comparing ML-GF outputs to DICOM Display images*

The ML-GF algorithm was also tested on raw DICOM images from UMPC Breast Tomography and FFDM Collection. These had ASG, so no scatter correction step was performed. The breast thickness information was available in DICOM. Where the paddle information was absent, we assumed 2.5mm paddle thickness. The skin thickness was assumed to be 1.45mm (as used in the literature) [3]. Effective energy was estimated using spectrum information in DICOM header (W anode with Ag filter 0.05mm). Where non-breast tissue was present (such as pectoral muscle) they were segmented out before applying the algorithm.

DICOM FFDM clinical images are often not available as raw detector counts. Instead, they undergo a series of processing steps for high contrast display. Initially, the log-ratio, $\ln(I_0/I_m)$ is calculated, followed by the application of various vendor-specific functions. To facilitate the GF algorithm, it is important to have access to the original raw images.

To conduct the micro-calcification analysis as well, raw images are required. Micro-calcifications are added as attenuated counts within the raw data. Afterward, it's necessary to estimate the vendor-specific transformation for displaying the images realistically for clinicians' assessment.

Therefore, we considered just the one DICOM case available where the raw linear counts and the processed display images were both available. Fortunately, this case had 4 different views of the breast, which effectively created different unique pixel-wise glandular fractions due to different view angles of projection. We applied the ML-GF algorithm for those cases using the linear image. We used this set for the micro-calcification analysis as well.

Since our ML-GF algorithm needed a $I_0$ we estimated that from outside the object column-wise. In the results we compare the ML-GF directly with the log-processed high contrast display images that were provided.

Four images were downloaded of the same breast: right craniocaudal (R CC), left craniocaudal (L CC), right mediolateral oblique (R MLO), and left mediolateral oblique (L MLO), and the DICOM images were ran through the same process the TOPAS images went through, (except for the Scatter-to-Primary estimate and scatter correction step because the clinical images were acquired with anti-scatter-grids). All images for this case were taken with a tungsten anode. This was the same spectrum that we used for the TOPAS Monte Carlo. The effective energy was estimated using the same method as the TOPAS images and by using a similar setup in TOPAS with the same parameters listed in the DICOM header.

*B. Adding Microcalcifications to DICOM images and comparing to ML-GF*

We added microcalcifications similar to TOPAS images. These microcalcifications were added as attenuation of raw counts in a denser region of interest in the densest sections of glandular tissue within the actual linear raw DICOM images. The attenuation was estimated from the attenuation by micro-calcifications for the TOPAS case. This is done at different spots for the four views for the single-patient for whom the raw data is available. Then the ML-GF algorithm was applied as in previous Section 2.6.1. We wish to test the visibility of microcalcifications on the ML-GF image. To ensure a fair display comparison, we need to compare ML-GF with micro-calcifications to a high-contrast DICOM version, not raw linear count images. This was done in a few steps explained below and shown in Fig 5.

1. Take the log-ratio $\ln(I_0/I_m)$ of the original raw count images (prior to micro-calcification addition) heretofore called Log-converted or Log-processed "linear" image Fig (5(a)).
2. Pixel-wise estimate the vendor mapping between this log-converted images Fig. 5(a) and the high-contrast DICOM display images available for this original set (Fig. 5(b))
3. Apply the log-conversion (Fig. 5(c)) and the estimated mapping in Step 2 above to the log-ratio images with micro-calcifications to obtain display images with calcifications (Fig. 5(d))

The calcification is hardly visible in these images -- in Results section we compare these with the ML-GF images with calcifications to see if visibility improved. The CNR at the locations of the microcalcifications was calculated and compared between the input and output images $CNR = \frac{x_s - x_{bg}}{\sigma_{bg}}$ where $x_s$ is the signal strength of calcification, $x_{bg}$ is the signal strength of the surrounding glandular tissue and $\sigma_{bg}$ is the standard deviation of the surrounding glandular tissue.

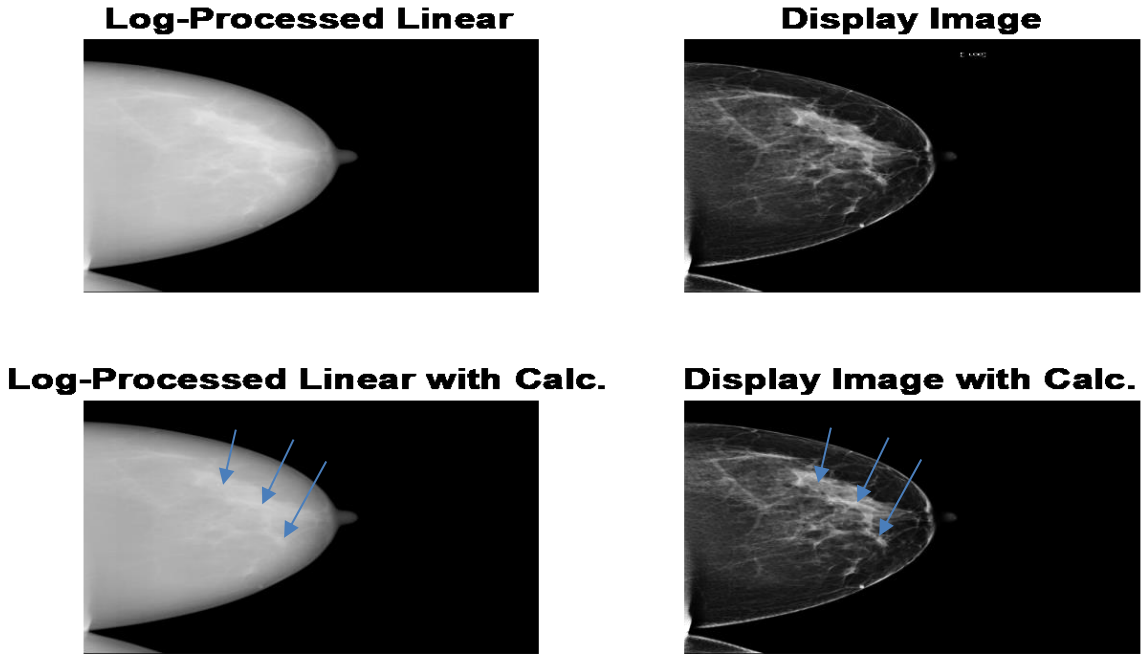

*Figure 5. Mapping is obtained from (Top Left) Log-converted Linear DICOM image to the (Top Right) log DICOM (Display) Image. This mapping is then applied to Log-Converted Linear image with micro-calcifications (Bottom Left) to obtain similar Display Image (Bottom Right), with calcification (arrows). Note: The calcification is hardly visible in these images.*

## 3. RESULTS

*Results for Topas Simulated Images*

In Fig. 6 the effective energy attenuation and the true measured poly-energetic attenuation are shown along with the error between the two in Table I. Both attenuation values are taken from TOPAS Monte Carlo simulations, which measured total counts with object and total counts without object. The effective energy values were estimated using linear interpolation of the linear attenuation coefficients. The errors between the poly-energetic attenuation and monoenergetic attenuation at the effective energy are less than 1% for all glandular fraction values tested. The low errors, in the results of Table I and Fig. 6(a) validated the approach.

We also tested the effective energy for different breast tissue thickness. The effective energies for the 3cm, 4cm, 5cm breasts ranged from 21keV to 21.5 keV for the 28kV W/Al spectrum. Fig. 6(b) shows the trend in the effective energy against breast thickness. First, when the total breast thickness is held constant and the glandular tissue thickness increases, the effective energy increases as seen in the in the red, black, and blue bars of Fig. 6(b). Second, as the total breast thickness increases, the effective energy also increases, as seen in the third red bar, which is 21 keV and the third blue bar, which is 21.5 keV. The attenuation results shown in Fig. 6(a) were used to calculate the effective energy.-These effective energy values are then used to get the $\mu_l, \mu_s, \mu_a, \mu_g$ values to be used in (6) from NIST XCOM.

In past literature, the SPR was calculated solely as a function of the total breast thickness, however other literature exists that states the SPR could change up to 15% depending on the glandular composition [27]. The SPR was calculated here using known scatter from the Monte Carlo histories Effective energy attenuation coefficients were utilized since the spectrum was poly-energetic.

Table I: Percent Errors of Images obtained with poly-energetic attenuation vs that with effective energy attenuation

|  | 3cm | 4cm | 5cm |
| --- | --- | --- | --- |
| 20% GF | 0.90% | 0.83% | 0.76% |
| 30% GF | 0.85% | 0.83% | 0.75% |
| 50% GF | 0.87% | 0.83% | 0.75% |

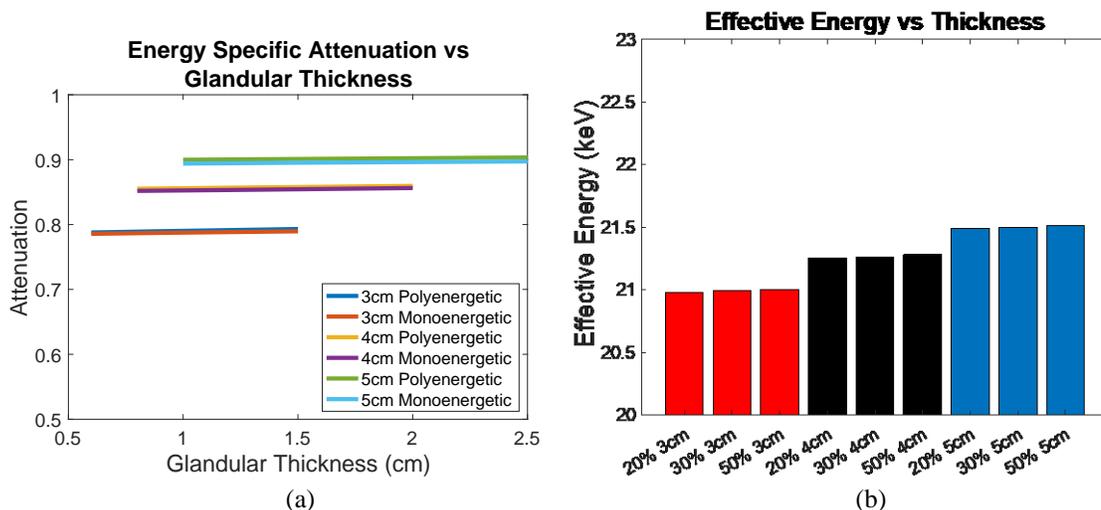

*Figure 6. (a) Poly-energetic and Monoenergetic (effective energy) Attenuation vs Glandular Thickness. The attenuation values were used to verify the effective energy of the W/Al spectrum, which is shown in (b) for each of different glandular thicknesses.*

We analyzed data for breast thicknesses of 3 cm, 4 cm, and 5 cm, considering glandular fractions of 20%, 30%, and 50%. The results from the two algorithms, **ML-ScRmv/ML-GF** and **ASG-ScRmv/ML-GF**, using images from TOPAS Monte Carlo simulations, are shown in Fig. 7 (a-b). This figure shows the average glandular fraction, standard deviation, and the true values of 20%, 30%, and 50%. Note, the average values fall within one to two standard deviations of the true value of glandular fraction.

The ML-ScRmv/ML-GF and ASG-ScRmv/ML-GF produced similar root mean square (RMSEs) of 3.2% and 2.5% deviations from the true GF value. Average absolute error were (3.7 ±2.4)% and (3.6 ± 0.9)% respectively.

We also investigated whether the location of the glandular tissue significantly impacts ML-GF estimation. Additional glandular z positions were considered beyond the middle of the breast. We moved the glandular tissue in a 4cm breast with 20% glandular fraction 1cm closer to the surface, in the middle and 1 cm away from the surface. The average values are 0.215 ± 0.045, 0.217 ± 0.045, and 0.220 ± 0.045 for middle, superior, and inferior respectively. These results show that there is no significant difference in glandular fraction estimation on z location of glandular tissue.

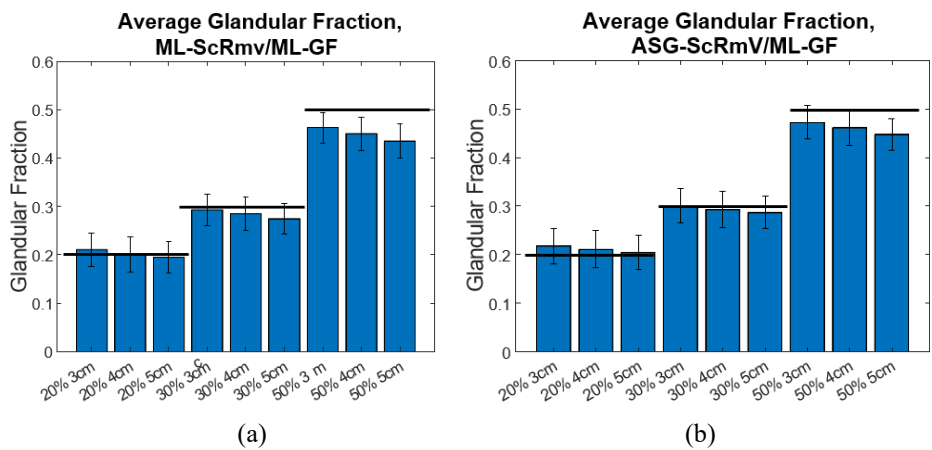

*Figure 7. Average Glandular Fraction with error-bars for (a) **ML-ScRmv/ML-GF** which corresponds to software scatter removal followed by glandular fraction estimate (b) **ASG-ScRmV/ML-GF** corresponds to anti-scatter grid removal technique followed by glandular fraction estimate. Black horizontal bars show the true value at each glandular fraction.*

*Results for Clinical Images and Microcalcification analysis:*

The breast thicknesses varied between 86mm and 98mm, for the DICOM images, depending on the view, and the peak energies of the spectrums were 33kVp, 34kVp, 35kVp, and 34kVp for R CC, L MLO, L CC, and R MLO respectively.

Figs 8 and 9 display DICOM images with added micro-calcifications (following Steps 1-3 outlined in the methods) and the corresponding output glandular fraction (GF) images generated by the ML-GF algorithm for all four views (the right column).

In these images, the GF varies, from 40% in the brightest regions to around 10% towards the chest wall. Since the true glandular fraction values are unknown, we cannot estimate error. However, the average of around 30% in the CC view aligns with the average glandular fraction range found in the literature [3].

The GF images for calcifications in Fig. 8-9, show some visibility improvement over the display DICOM images. Note no special processing was applied on the GF images. While visibility is still subtle because the macrocalcifications were very small, (1mm diameter) there was a CNR improvement of nearly 550% in the L MLO (mediolateral oblique view) while the lowest improvement was a 17.5% improvement in one of the microcalcifications' in the same L MLO view. Fig. 10(a) shows the contrast-to-noise ratio on all the three microcalcifications that were added in each image, for a total of 12 separate calcifications across the images. Fig. 10(b) also show improvement in contrast-to-noise for microcalcifications for Topas images for 50% GF and 30% GF images. These were computed with respect to unprocessed (raw count) images.

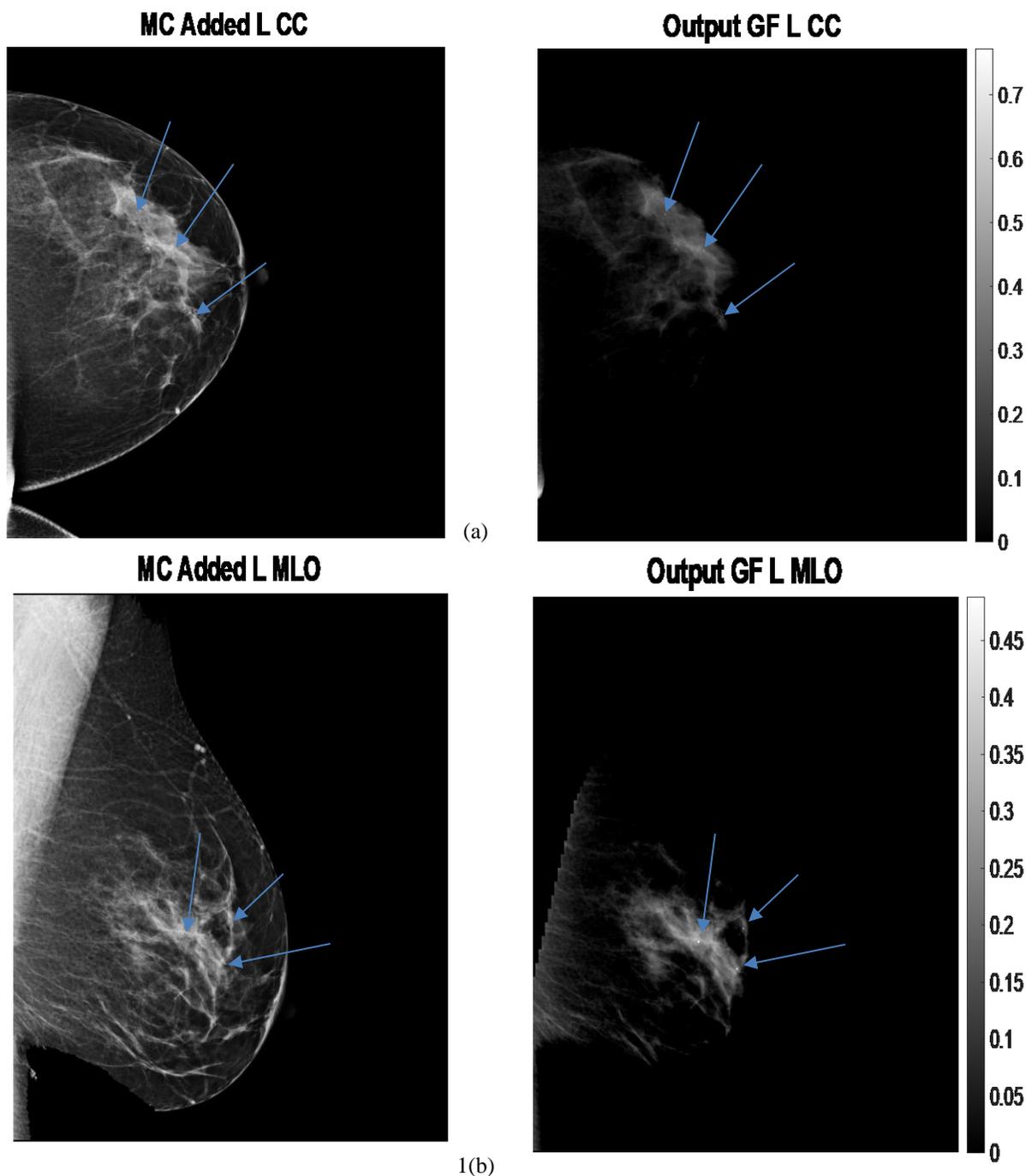

*Figure 8. (Left) Microcalcification (MC) added DICOM images and (Right) Output Glandular Fraction (GF) Images (output by algorithm). Different views are (a) left craniocaudal view, (b) left medio-lateral oblique view. Color bars in right images represent glandular fraction. Note the MCs that are virtually invisible in DICOM images become noticeable in GF-images.*

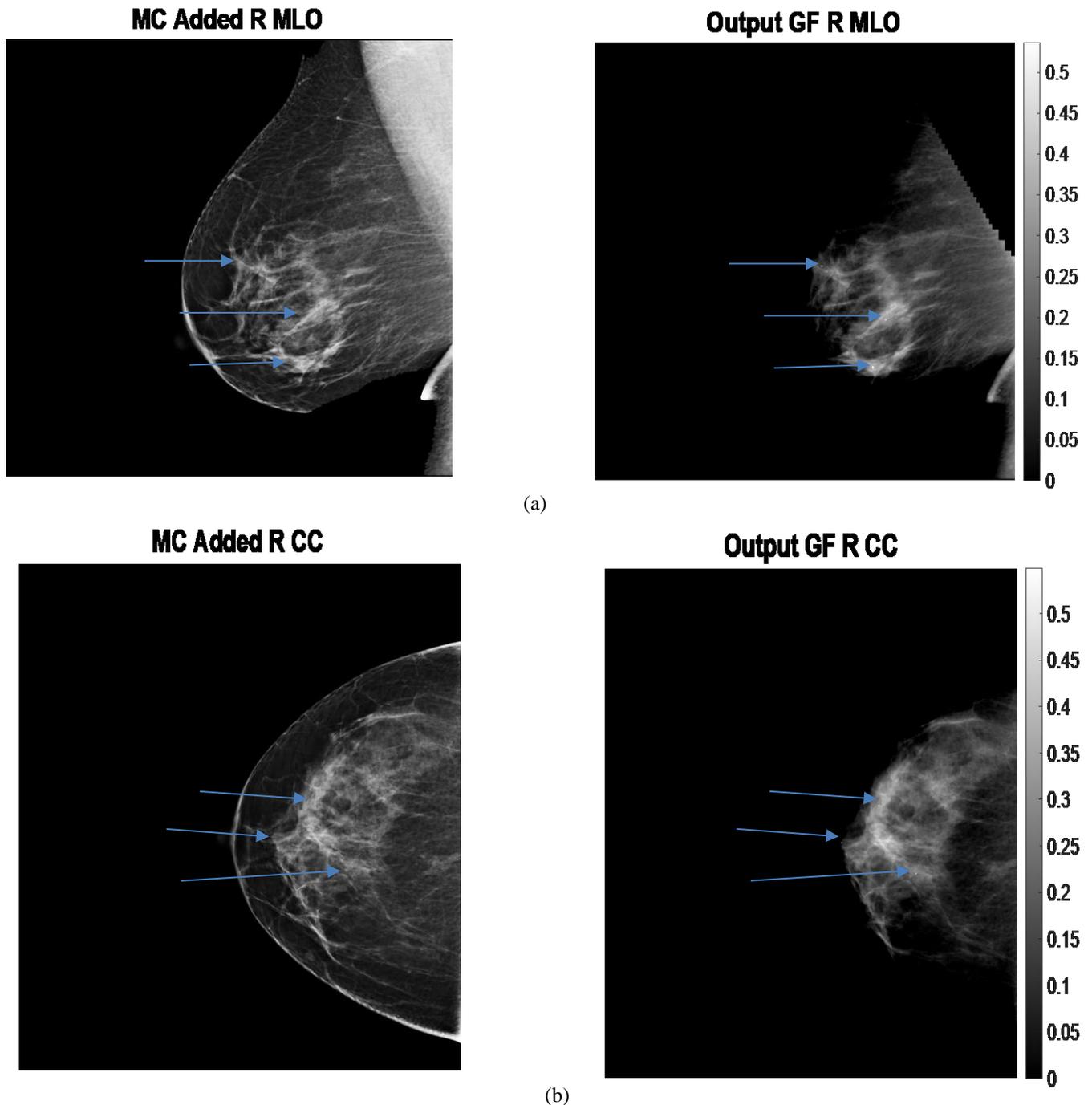

*Figure 9. (Left) Microcalcification (MC) added DICOM images and (Right) Output Glandular Fraction (GF) Images (output by algorithm). Different views (a) right mediolateral oblique view, and (b) right craniocaudal view of 9cm breast. Color bars in right images represent glandular fraction. Note the MCs that are virtually invisible in DICOM images become noticeable in GF-images.*

## 4.DISCUSSION

The results of the TOPAS simulations show that glandular fraction can be estimated within an error of 3.2% where the software-based scatter removal was used prior to GF estimate (**ML-ScRmv/ML-GF**) and 2.5% for the anti-scatter grid scatter removal prior to GF estimate (**ASG-ScRmv/ML-GF**). Particularly, the results for the **ASG-ScRmv/ML-GF** case show a more accurate GF estimate at the 50% case when compared to the **ML-ScRmv/ML-GF** at 50% glandular fraction. Overall, the ASG case leads to higher average GF estimates on each setup.

Simulations are robust to different parameters. The focal spot blur is not explicitly simulated but its effect is minimal since the object is next to the detector, making the magnification small. For example, for source-detector distance 70cm, the worst case (for top surface of breast) effect of the focal spot blur of 300μm on the detector for a 4 cm breast with a 1.5cm gap to the detector will be 5.5/(70-5.5) x 300μm = 25.6μm, which is subpixel for a 50 μm flat panel detector. Smaller detector thickness also will not have a significant effect on the GF estimation in terms of detection sensitivity, because our count levels (120 million x 8 energy bins) are several orders lower than true clinical count levels.

This work shows that an absolute physics model can be used to estimate GF to absolute accuracy of within 4% for worst case scenario. Limitations of this work is the $I_0$ that is needed in the estimation of the GF, which may need access to raw data. Previous literature does not need an $I_0$ value due to the relative physics model, but this work shows that an absolute model is necessary to yield quantitatively accurate GF. This fact restricted our clinical data application to one case where raw linear images were available. We could estimate the $I_0$ from the background as well. Fortunately, the availability of 4 different views of the same patient led to cases of different overlaps of tissue and created unique pixelwise GF for us to evaluate.

While the data for clinical DICOM images has no truth to compare, qualitatively the pixel-wise GF corresponded to excellent segmentation of glandular tissue versus background as shown in Fig. 8 and 9. The GF values obtained were also well within the expected values clinically. Some data required by the complete algorithm maybe absent such as the paddle thickness. In such cases one can attempt to find if the specific information is available for the system described in the DICOM header. If that fails, we can assume standard values in the literature for the system, which should be close. Errors in paddle thickness adds a small scaling error in the estimates, particularly for smaller breast thickness. Last but not the least, the ML-GF images seem to make visible realistic small micro-calcifications that are nearly invisible in displayed clinical DICOM images. The CNR improved on average 200% for clinical images. This makes it a potentially viable image to show physicians in addition to the the high contrast DICOM images. Lastly, the methods can be directly applied to other advanced applications, such as attenuation image for a X-ray interferometry system.

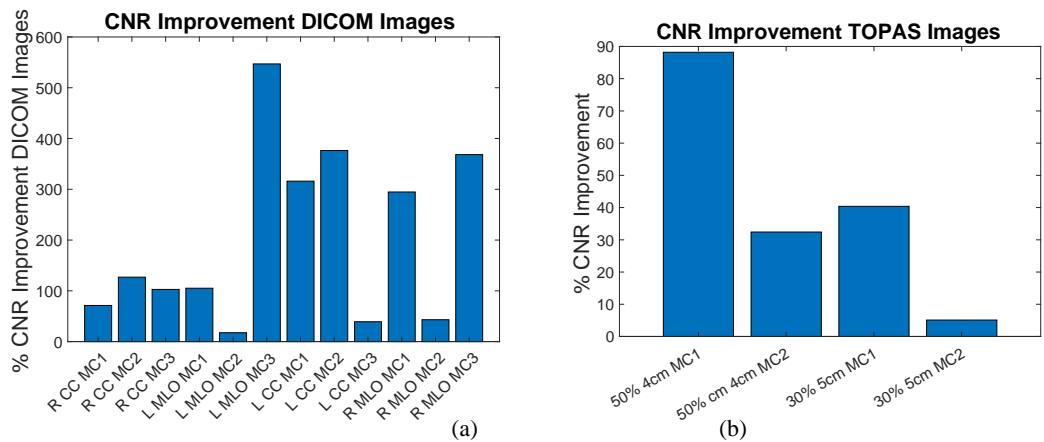

Figure 10. Percent improvement on contrast-to-noise ratio for Microcalcifications (a) on GF estimated images from DICOM images (b) on the GF-images outcomes for TOPAS images.

## 5. CONCLUSION

We developed a ML algorithm to estimate the pixel wise glandular fraction of an object. We evaluated the ML software with images generated from TOPAS Monte Carlo simulations and our estimated glandular fraction converges to the true glandular fraction with an RMSE of 3.2% and 2.5% for software-based and anti-scatter grid respectively for 3cm, 4cm, and 5cm breast thicknesses. Average absolute error were (3.7 ±2.4)% and (3.6 ± 0.9)% respectively. Furthermore, this maximum-likelihood estimation method showed an avg of 200% improvement in contrast-to-noise on microcalcification for DICOM images. The realistic micro-calcifications were nearly invisible in original displayed imaged but were visible in the glandular fraction images. This is further important because there has been shown to be a proportional relationship between glandular fraction and microcalcifications and breast cancer risk.

**ACKNOWLEDGEMENTS**
The work presented here is in part from first-author Bryce Smith's MSc thesis work (advised by co-author J. D), while he was a graduate student at LSU [28] Department of Physics and Astronomy, Medical Physics program, LSU, Baton Rouge, LA. This work was supported by NIH NIBIB Trail-blazer Award 1-R21-EB029026-01A1.